\documentclass[%
 reprint,
showpacs,preprintnumbers,
 amsmath,amssymb,
 aps,
]{revtex4-1}

\usepackage{graphicx}
\usepackage{multirow}

\newcommand{\myfig}[4][ht]{
\begin{figure}[#1]
\centering
\includegraphics[#2]{#3}
\caption{#4\label{#3}}
\end{figure}
}

\newcommand{\myfigwide}[4][ht]{
\begin{figure*}[#1]
\centering
\includegraphics[#2]{#3}
\caption{#4\label{#3}}
\end{figure*}
}

\newcommand{\npar}{%
\\[-1mm]\par}
\graphicspath{{images/}}

\usepackage{dcolumn}
\usepackage{bm}

\begin{document}


\title{Cross-plane heat conduction in thin films with\\ab-initio phonon dispersions and scattering rates}
\author{Bjorn Vermeersch}
\email{bjorn.vermeersch@cea.fr}
\author{Jes\'{u}s Carrete}
\author{Natalio Mingo}
\affiliation{\vspace{3mm}CEA, LITEN, 17 Rue des Martyrs, 38054 Grenoble, France}
\date{\today}
\begin{abstract}%
We present a first-principles study of the cross-plane thermal conductivity $\kappa_{\perp}$ in a wide variety of semiconductor thin films. We introduce a simple suppression model that matches variance-reduced Monte Carlo simulations with \textit{ab-initio} phonon dispersions and scattering rates within $\leq 5\%$ even for anisotropic compounds. This, in turn, enables accurate $\kappa_{\perp}$ reconstruction from tabulated cumulative conductivity curves $\kappa_{\Sigma}(\Lambda_{\perp})$. We furthermore reveal, and explain, a distinct quasiballistic regime characterised by a fractional thickness dependence $\kappa_{\perp} \sim L^{2-\alpha}$ in alloys (where $\alpha$ is the L\'evy exponent) and logarithmic dependence $\kappa_{\perp} \sim \ln(L)$ in single crystals. These observations culminate in the formulation of two compact parametric forms for $\kappa_{\perp}(L)$ that can fit the first-principles curves across the entire ballistic-diffusive range within a few percent for all investigated compounds.%
\end{abstract}
\maketitle
Phonon-mediated heat conduction in thin films plays an important role in nanoscale devices \cite{cahillreview1,cahillreview2} and received increasing theoretical attention \cite{majumdar,extraref1,extraref2,extraref3,minnichthinfilm,SOIfilms,nitrides,kappacumulcurves}. Interestingly, cross-plane thermal transport has posed a considerably harder challenge than its in-plane counterpart. An exact solution of the Boltzmann transport equation (BTE) in cross-plane geometries has been obtained only very recently \cite{minnichthinfilm} and provides the film conductivity at thickness $L$ as an integral over phonon frequency $\kappa(L) = \int S(\omega,L) \, \kappa(\omega) \, \mathrm{d}\omega$. Here $S$ is a `suppression function' that contains the thin film physics and $\kappa(\omega)$ denotes the bulk spectral conductivity. Evaluation of the latter often relies on analytical models with isotropic dispersions \cite{SOIfilms,nitrides,kappacumulcurves} for mathematical convenience. More realistic phonon spectra can also be utilised through frequency binning, but this procedure becomes problematic in anisotropic compounds. In this work, we adopt a first-principles framework \cite{ShengBTE_2014,abinitiobookchapter} that goes well beyond the predictive power of spectral formulations. We subsequently show that the resulting cross-plane thin film conductivity curves $\kappa(L)$ can be captured by compact models that effortlessly maintain their accuracy in anisotropic compounds. In the process, we also reveal quasiballistic regimes in both alloys and single crystals with distinct film thickness dependences.
\npar
We performed our \textit{ab-initio} study for a variety of semiconductors selected for their technological relevance. Thin Si films play a key role in silicon-on-insulator devices \cite{SOIfilms}; nanostructured SiGe alloys are of interest to thermoelectric systems \cite{sige}; InAs and InGaAs serve as basis for high electron mobility transistors \cite{arsenides} while III-V nitrides such as GaN and AlGaN find application in high-power transistors and UV optoelectronics \cite{nitrides}. SnSe received recent attention for its thermoelectric properties \cite{snsenature}. Given its triaxial phononic anisotropy \cite{snse}, it additionally provides a challenging benchmark test for the thin film conduction model we present below.
\par%
We calculated \textit{ab-initio} heat capacities $C(\vec{k})$, group velocities $\vec{v}(\vec{k})$ and scattering rates $\tau^{-1}(\vec{k})$ using the procedure described in Ref. \onlinecite{levy1}. The adopted method has been tested on a variety of bulk compounds and provides results in good agreement with experiments \cite{ShengBTE_2014}. Phonon properties are resolved over a uniform wavevector grid with $N_k^3$ points, where we have set $N_k$ to 24 for diamond/zincblende crystals, 16 for wurtzites and 12 for Pnma SnSe. Scattering rates $\tau^{-1} = \tau_{\text{anh}}^{-1} + \tau_{\text{har}}^{-1}$ and mean free paths (MFPs) $\Lambda = \| \vec{v} \| \tau$ account for anharmonic (three-phonon) and harmonic (isotope/alloy) scattering mechanisms. Alloys were treated under the virtual crystal approximation \cite{MgSiSn,katcho_lattice_2012, economou}. Further technical details on the \textit{ab-initio} framework are available elsewhere \cite{ShengBTE_2014,abinitiobookchapter}. Table I summarises computed values for bulk properties relevant to this study, including the thermal conductivity
\begin{equation}
\kappa_{\text{bulk}} = \sum \kappa(\vec{k}) = \sum C(\vec{k}) \, v_{\perp}(\vec{k}) \, \Lambda_{\perp}(\vec{k})
\end{equation}
where $\perp$ signals cross-plane projection $X_{\perp} = X \, |\cos \theta|$ with $\theta$ the group velocity angle to the film surface normal. Summations span all wavevector points in the Brillouin zone (BZ) and implicitly run over polarisation/branch index. The table also lists the `dominant cross-plane MFP' which we defined as the following weighted average
\begin{equation}
\Lambda_{\text{dom}} = \sum \frac{\kappa(\vec{k})}{\kappa_{\text{bulk}}} \cdot \Lambda_{\perp}(\vec{k}) \label{dominantMFP}
\end{equation}
This metric conveys a characteristic MFP of the modes that govern most of the heat conduction and thus sets a film thickness threshold below which we should expect notable deviations from bulk thermal transport.%
\begin{table}[!htb]%
\caption{\textit{Ab-initio} bulk conductivity, dominant cross-plane MFP (see Eq. \ref{dominantMFP}) and alloy L\'evy exponent $\alpha$ at 300$\,$K. $L_{\text{b}}$ and $L_{\text{d}}$ are best fitting characteristic length scales for parametric film conductivity formulae (\ref{parametricalloy}) and (\ref{parametricsinglecrystal}).}
\vspace{3mm}
\begin{tabular}{cccccc}
\hline\hline
\multirow{2}{*}{Compound} & $\kappa_{\text{bulk}}$ & $\Lambda_{\text{dom}}$ & \multirow{2}{*}{$\alpha$} & $L_{\text{b}}$ & $L_{\text{d}}$\\
 & [W/m-K] & [$\mu$m] & & [nm] & [$\mu$m]\\
\hline\hline
\multicolumn{6}{c}{\textit{Cubic/diamond crystals (`isotropic')}}\\
\hline
Si & 156 & 2.59 & --- & 17.6 & 32.5\\
Ge & 53.3 & 0.96 & --- & 13.9 & 10.7\\
Si$_{0.99}$Ge$_{0.01}$ & 42.2 & 2.99 & 1.699 & 3.4 & 16.4\\
Si$_{0.82}$Ge$_{0.18}$ & 7.56 & 1.04 & 1.641 & n/a & 4.2\\
\hline
\multicolumn{6}{c}{\textit{Cubic/zincblende crystals (`isotropic')}}\\
\hline
InAs & 33.9 & 1.60 & --- & 26.4 & 22.5\\
GaAs & 42.7 & 0.37 & --- & 18.0 & 3.8\\
In$_{0.53}$Ga$_{0.47}$As & 8.28 & 0.72 & 1.694 & 2.7 & 3.9\\
\hline
\multicolumn{6}{c}{\textit{Wurtzite crystals (a \& b axes)}}\\
\hline
GaN & 203 & 0.73 & --- & 40.6 & 6.6\\
AlN & 284 & 0.44 & --- & 35.3 & 3.6\\
Al$_{0.01}$Ga$_{0.99}$N & 82.1 & 1.00 & 1.715 & 12.0 & 5.8\\
Al$_{0.1}$Ga$_{0.9}$N & 24.4 & 0.69 & 1.724 & 2.8 & 4.9\\
Al$_{0.4}$Ga$_{0.6}$N & 9.53 & 0.36 & 1.722 & 1.1 & 2.6\\
\hline
\multicolumn{6}{c}{\textit{Wurtzite crystals (c axis)}}\\
\hline
GaN & 183 & 1.57 & --- & 23.6 & 15.7\\
AlN & 278 & 0.78 & --- & 22.5 & 4.9\\
Al$_{0.01}$Ga$_{0.99}$N & 99.7 & 2.40 & 1.720 & 10.4 & 12.8\\
Al$_{0.1}$Ga$_{0.9}$N & 38.7 & 2.75 & 1.724 & 2.9 & 17.3\\
Al$_{0.4}$Ga$_{0.6}$N & 18.8 & 2.73 & 1.738 & 1.5 & 22.9\\
\hline
\multicolumn{6}{c}{\textit{Pnma crystal}}\\
\hline
SnSe (\textit{a} axis) & 0.59 & 0.10 & --- & 0.39 & 0.74\\
SnSe (\textit{b} axis) & 1.60 & 0.010 & --- & 1.60 & 0.078\\
SnSe (\textit{c} axis) & 0.81 & 0.005 & --- & 1.10 & 0.033\\
\hline\hline
\end{tabular}
\vspace{-3mm}
\end{table}%
\par%
The \textit{ab-initio} phonon properties serve as basis for variance-reduced Monte Carlo (VRMC) simulations \cite{VRMC} of the energy-based RTA-BTE $\partial_t \, g_k + v_{k,x} \, \partial_x \, g_k = -(g_k - C_k \Delta T)/\tau_k$. We consider a slab structure with thickness $L$ and fully absorbing, isothermal walls at temperatures $T_{\text{ref}} \pm \Delta T/2 = 300 \pm 0.1\,$K. Dividing the steady state heat flux (easily evaluated from particle ensemble averaging) by the thermal gradient $\Delta T / L$ yields the apparent cross-plane conductivity $\kappa(L)$. We consider thicknesses down to 1$\,$nm so as to verify ballistic asymptotics, though we note that the BTE itself will lose validity at such extremely small scales. For each considered thickness, we performed 3 independent simulation runs with one million particles each and used the average result. Tests with 30 runs produced near-identical averages; the stochastic uncertainty on the reported values amounts to $\leq 0.5\%$. Results for cubic crystals pertain to the (001) principal axis, but we have verified that $\kappa(L)$ curves even for (111)-oriented films remain within 4\% of the reported data. Cubic compounds can thus be considered as virtually isotropic for cross-plane conduction purposes.
\par
Figure \ref{fig1-kappafilm} shows the obtained film conductivities. Two well established limit regimes are easily recognised. In very thin films (in the sense that $L \ll \Lambda_{\text{dom}}$) heat conduction is predominantly ballistic, and we indeed observe the familiar linear dependence $\kappa(L) \sim L$. Thick films ($L \gg \Lambda_{\text{dom}}$), by contrast, operate in the quasi-diffusive limit $\kappa(L) \simeq \kappa_{\text{bulk}} \sim L^0$. At intermediate thicknesses $L \simeq \Lambda_{\text{dom}}$, alloys display a distinct fractional dependence $\kappa(L) \sim L^{\gamma}$ ($0 < \gamma < 1$). Far more than a transitionary feature, this regime persists over 2--3 orders of magnitude of film thickness and appears even at 1\% alloying. Although seemingly minute, such concentrations exceed typical dopant and intrinsic impurity levels by several orders of magnitude, and they do have a notable effect on the phonon scattering rates (Fig. \ref{fig2-scatteringrates}). As we will explain shortly, it is precisely the phonon scattering relation that regulates the quasiballistic portion of the $\kappa(L)$ curve and instigates the fractional thickness dependence in alloys.
\myfigwide[!htb]{width=0.82\textwidth}{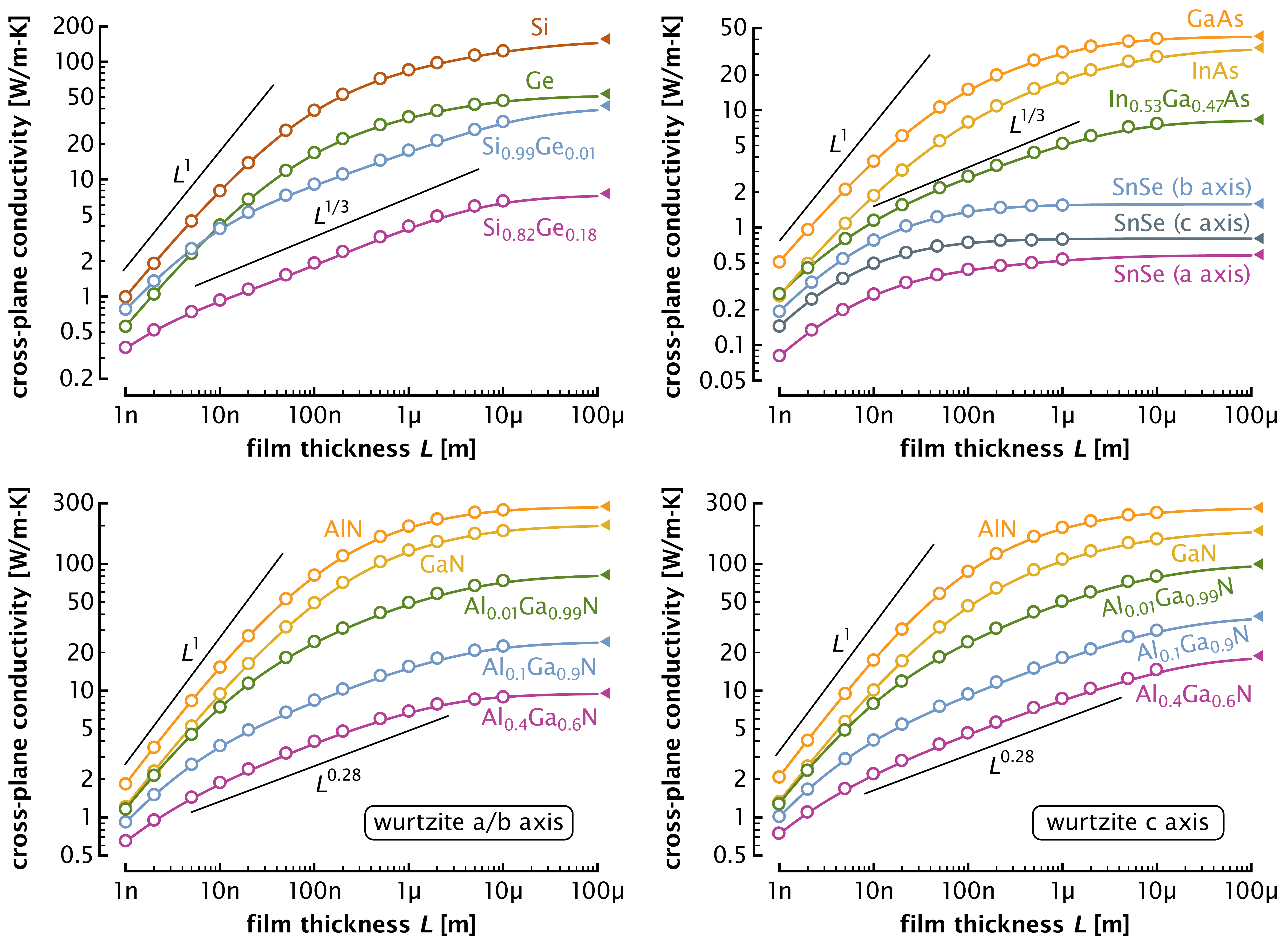}{(Color online.) Cross-plane thermal conductivity in thin films computed from \textit{ab-initio} phonon dispersions and scattering rates. Symbols ($\circ$) denote Monte Carlo solutions, curves (---) depict model predictions per Eq. (5), and triangles ($\blacktriangleleft$) mark the bulk conductivities. A fractional thickness dependence is distinctly visible for the alloy compounds.}%
\myfig[!htb]{width=0.42\textwidth}{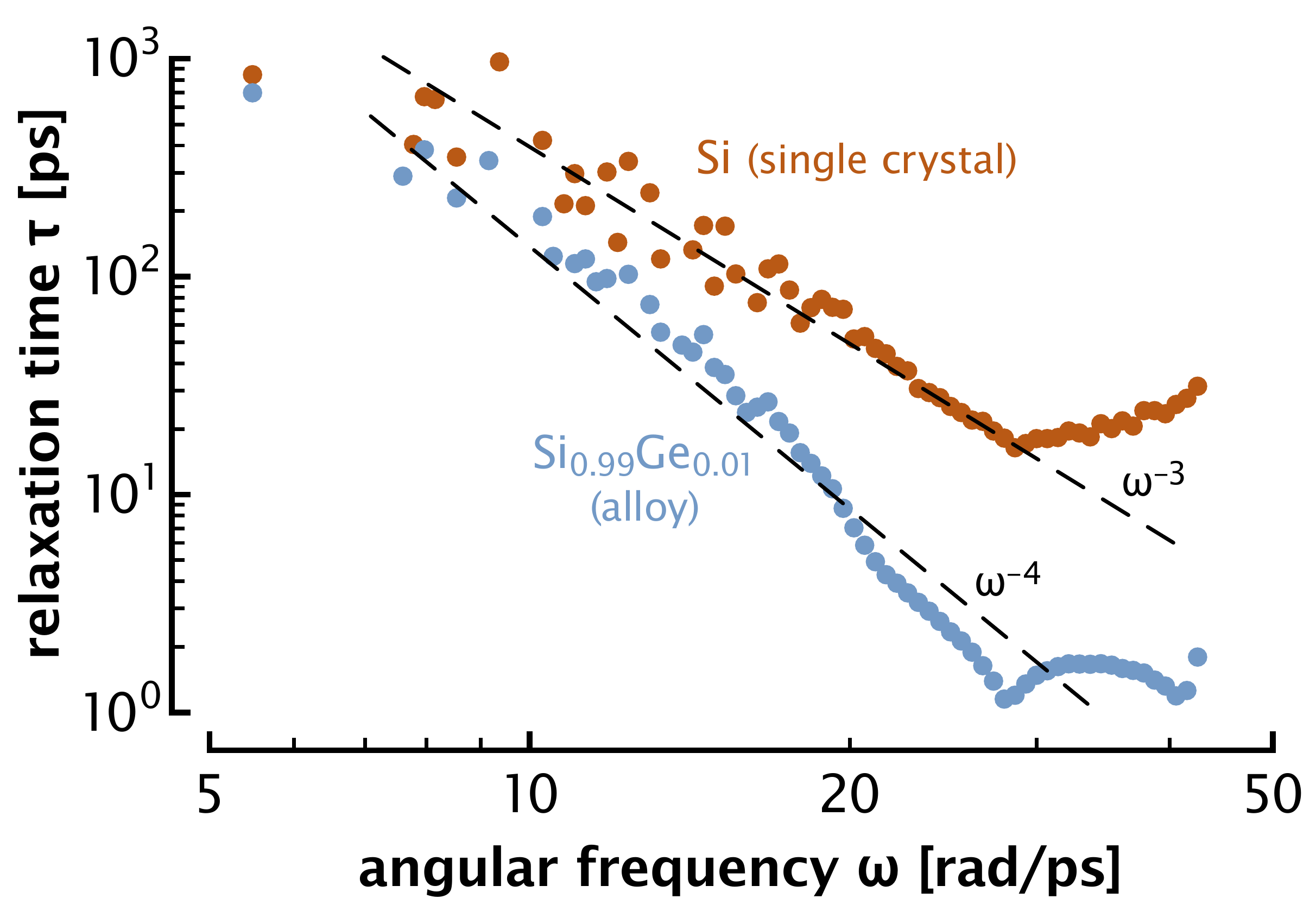}{Wavevector-averaged \textit{ab-initio} phonon relaxation times of the TA branches in Si and Si$_{0.99}$Ge$_{0.01}$.}
\npar%
Contrary to spectral formulations, the full \textit{ab-initio} phonon properties we have used in this work preserve the fine details of preferential heat propagation directions. The question now arises whether we can describe cross-plane conductivity by a \textit{modal} suppression model
\begin{equation}
\kappa(L) = \sum S(\vec{k},L) \, C(\vec{k}) \, v_{\perp}(\vec{k}) \Lambda_{\perp}(\vec{k}) \label{basicformula}
\end{equation}
Even in anisotropic compounds, time reversal symmetry of the lattice wave equations assures that phonons at wavevectors $\pm \vec{k}$ have identical frequencies and opposite group velocities. Each of these mode pairs contribute a cross-plane conductivity $\kappa_{\perp}(\vec{k},L) = \mathrm{Tr}(\vec{k},L) \, C(\vec{k}) \, v_{\perp}(\vec{k}) \, L$. The cross-plane transmission coefficient $\mathrm{Tr}$ can only depend on the `projected' Knudsen number $K_{\perp} = \Lambda_{\perp}/L$ due to geometric arguments. Compared to the bulk conductivity contribution $2 \, C(\vec{k}) \, v_{\perp}(\vec{k}) \, \Lambda_{\perp}(\vec{k})$, the pair is effectively suppressed by a factor $S(\vec{k},L) = \mathrm{Tr}(K_{\perp})/(2K_{\perp})$. Random walk simulations show that $\mathrm{Tr} = 2K_{\perp}/(1 + 2K_{\perp})$ and hence
\begin{equation}
S(\vec{k},L) = \frac{1}{1 + 2 \Lambda_{\perp} / L} = \frac{1}{1 + 2 K |\cos \theta|} \label{suppressionfunction}
\end{equation}
We can thus postulate, without actually solving the BTE,
\begin{equation}
\kappa(L) \simeq \sum \limits \frac{C(\vec{k}) \, v_{\perp}(\vec{k}) \Lambda_{\perp}(\vec{k})}{1 + 2 \Lambda_{\perp}(\vec{k})/L} \label{kappaL}
\end{equation}
One could accordingly say that inside the film, each mode possesses an effective cross-plane MFP that obeys
\begin{equation}
\frac{1}{\Lambda_{\perp}^{\text{eff}}} = \frac{1}{\Lambda_{\perp}^{\text{bulk}}} + \frac{2}{L} \label{MFPreduction}
\end{equation}
Despite its simplicity, the model (\ref{kappaL}) matches the VRMC results within 5\% or better (Fig. \ref{fig1-kappafilm}) even in notably anisotropic compounds such as SnSe or Al$_{0.4}$Ga$_{0.6}$N.
\par%
For purposes of comparison with prior work, we also derive the isotropic spectral suppression of our model. The spectral conductivity in both film and bulk configurations is easily found through integrations over a spherical BZ shell. After eliminating common factors from the ratio $\kappa_{\text{film}}(\omega)/\kappa_{\text{bulk}}(\omega)$ we find
\begin{eqnarray}
S_{\omega}(K) & = & \int \limits_{0}^{\pi} \frac{\cos^2 \theta \, \sin \theta \, \mathrm{d}\theta}{1 + 2 K |\cos \theta|} \,\, \biggr / \int \limits_{0}^{\pi} \cos^2 \theta \, \sin \theta \, \mathrm{d}\theta \nonumber \\
& = & \frac{3}{8 K^3} \left[ 2K(K-1) + \ln(1+2K) \right] \label{spectralsuppression}
\end{eqnarray}
This function deviates nowhere more than 12\% from Hua and Minnich's result $S_{\omega}^{\text{HM}}(K) = 1 + 3K \left[ E_5 \left( 1/K \right) - 1/4 \right]$ for the semi-analytical exact solution of the BTE inside a finite domain \cite{minnichthinfilm}, and also matches the grey-medium formula $S = 1/(1+4K/3)$ introduced by Majumdar \cite{majumdar} within 4\%. All three spectral solutions exhibit identical ballistic asymptotes $S_{\omega}(K \gg 1) \simeq 3/(4K)$.
\par
In analogy to previous derivations for in-plane transport \cite{kappacumulcurves}, our findings allow us to establish a direct relation between the cross-plane film conductivity $\kappa(L)$ and the so called MFP spectrum $\kappa_{\Sigma}$, provided the latter be resolved for \textit{projected} MFP i.e.
\begin{equation}
\kappa_{\Sigma}(\Lambda_{\perp}^{\ast}) \equiv \sum \limits_{\Lambda_{\perp} \leq \Lambda_{\perp}^{\ast}} C(\vec{k}) \, v_{\perp}(\vec{k}) \, \Lambda_{\perp}(\vec{k})
\end{equation}
The practical utility of such a connection lies in the fact that MFP spectra are a well established concept whose interpretation does not require any knowledge on the underlying \textit{ab-initio} data and can be easily shared among researchers in tabulated form $\{ \kappa_{\Sigma}(\Lambda_{\perp}^{(m)}) \,\, | \,\, m = 1 \ldots M \}$. In a thin film, the bulk conductivity $\kappa_{\Sigma}(\Lambda_{\perp}^{(m+1)}) - \kappa_{\Sigma}(\Lambda_{\perp}^{(m)})$ contributed by the phonons with projected MFPs between $\Lambda_{\perp}^{(m)}$ and $\Lambda_{\perp}^{(m+1)}$ is approximately suppressed by the factor $S(\bar{\Lambda}_{\perp}^{(m)}/L)$ of the interval midpoint $\bar{\Lambda}_{\perp}^{(m)} = ({\Lambda}_{\perp}^{(m)} + {\Lambda}_{\perp}^{(m+1)})/2$ and so we have
\begin{equation}
\kappa(L) \simeq \sum \limits_{m=1}^{M-1} \frac{\kappa_{\Sigma}(\Lambda_{\perp}^{(m+1)}) - \kappa_{\Sigma}(\Lambda_{\perp}^{(m)})}{1 + (\Lambda_{\perp}^{(m)}+\Lambda_{\perp}^{(m+1)})/L}
\end{equation}
We have found that $M=250$ is sufficient to reproduce the actual film conductivities within $\leq$1\%. We are providing tabulated $\kappa_{\Sigma}(\Lambda_{\perp})$ functions for all investigated compounds as Supplemental Material \cite{supplmat}.%
\par%
We now return to the fractional length dependence in alloys. This effect, as well as the overall form of the $\kappa(L)$ curve in both alloys and single crystals, can be explained with a simple reasoning as follows. Let us consider a single isotropic Debye branch in high temperature regime ($C(k)$ is constant) with a scattering relation $\tau \sim \omega^{-n} \Leftrightarrow \Lambda(k) = A/k^n$ ($n$ not necessarily integer). Applying our suppression model (\ref{kappaL}) to the idealised crystal gives
\begin{equation}
\kappa(L) \sim \int \limits_{0}^{\pi} \cos^2 \theta \, \sin \theta \, \mathrm{d}\theta \int \limits_{k_{\text{min}}}^{k_{\text{max}}} \frac{k^2 \, \Lambda(k) \mathrm{d}k}{1 + 2 \Lambda(k) |\cos \theta| / L} \label{kappa_alloymodel}
\end{equation}
Integration bounds $k_{\text{min}}>0$ and $k_{\text{max}} < \infty$ physically arise because the finite size and nonzero lattice constant of real-world crystals enforce a maximum and minimum permissable phonon wavelength respectively. The Debye crystal supports an according MFP range that spans from $\Lambda_{\text{min}} \equiv A/k_{\text{max}}^n$ to $\Lambda_{\text{max}} \equiv A/k_{\text{min}}^n$. Replacing $|\cos \theta|$ by its average value $1/2$ for simplicity and changing the integration variable to $u = (\Lambda_{\text{max}}/L) \cdot (k_{\text{min}}/k)^n$ yields
\begin{equation}
\kappa(L) \sim L^{1-3/n} \, \int \limits_{\Lambda_{\text{min}}/L}^{\Lambda_{\text{max}}/L} \frac{\mathrm{d}u}{u^{3/n} \, (1 + u)} \label{kappaintegral}
\end{equation}
How this result behaves clearly depends on the sign of $1-3/n$. Let us first consider $n>3$. In diffusive regime $L \gg \Lambda_{\text{max}}$, the integral only spans values $u \ll 1$. The integrand behaves as $u^{-3/n}$ and thus produces a primitive function $\propto u^{1-3/n}$. With both integration bounds $\sim L^{-1}$, we find $\kappa(L) \sim L^0$ as appropriate. The ballistic regime $L \ll \Lambda_{\text{min}}$ deals with $u \gg 1$, where the integrand behaves as $u^{-3/n-1}$. The primitive function scales as $u^{-3/n}$ so we correctly recover $\kappa(L) \sim L^1$. In quasiballistic regime $\Lambda_{\text{min}} \ll L \ll \Lambda_{\text{max}}$, the integral only deviates from its extended counterpart $\int_{0}^{\infty} \mathrm{d}u/[u^{3/n} \, (1+u)] \,  = \pi/\sin(3\pi/n)$ by a negligable amount $\mathcal{O}[(\Lambda_{\text{min}}/L)^{1-3/n} + (\Lambda_{\text{max}}/L)^{-3/n}]$, and a fractional length dependence thus emerges:
\begin{equation}
n>3 : \quad \kappa(L) \sim L^{1-3/n} \sim L^{2-\alpha}
\end{equation}
Interestingly, this (steady state) result can be expressed in terms of the L\'evy exponent $\alpha = 1+3/n$ that plays a crucial role in the alloy's (transient) single pulse response \cite{levy1,levy2}. Ideal Reyleigh scattering ($n=4$) corresponds to $\alpha = 7/4$ and produces the $\kappa \sim L^{1/4}$ trend previously reported in molecular dynamics simulations \cite{fractionallaw}. It also offers a good approximation for the AlGaN results presented here ($\alpha \simeq 1.72$). SiGe and InGaAs compounds exhibit slightly lower L\'evy exponents $\alpha \simeq 5/3$ (see Table I above and experimental values in Ref. \onlinecite{levy2}) and thus induce a $\kappa(L) \sim L^{1/3}$ tendency. The observation that the $\kappa(L)$ curve in alloys performs its ballistic-diffusive transition through a power law with exponent $2-\alpha$ enables us to propose the following parametric form
\begin{equation}
\text{alloys: } \kappa(L) \simeq \frac{\kappa_{\text{bulk}}}{(1 + L_{\text{d}}/L)^{2-\alpha} \cdot (1 + L_{\text{b}}/L)^{\alpha-1}} \vspace{2mm} \label{parametricalloy}
\end{equation}
where $L_\text{b}$ and $L_{\text{d}}$ are two characteristic film thicknesses that demarcate the transition to ballistic and diffusive regimes respectively. Although the Debye reasoning we adopted is an obvious oversimplification of realistic crystals, the resulting functional form (\ref{parametricalloy}) can fit the \textit{ab-initio} model curves (\ref{kappaL}) with remarkable accuracy for all investigated alloy compounds (Fig. \ref{fig3-parametriccurves1}).%
\myfigwide[!htb]{width=0.85\textwidth}{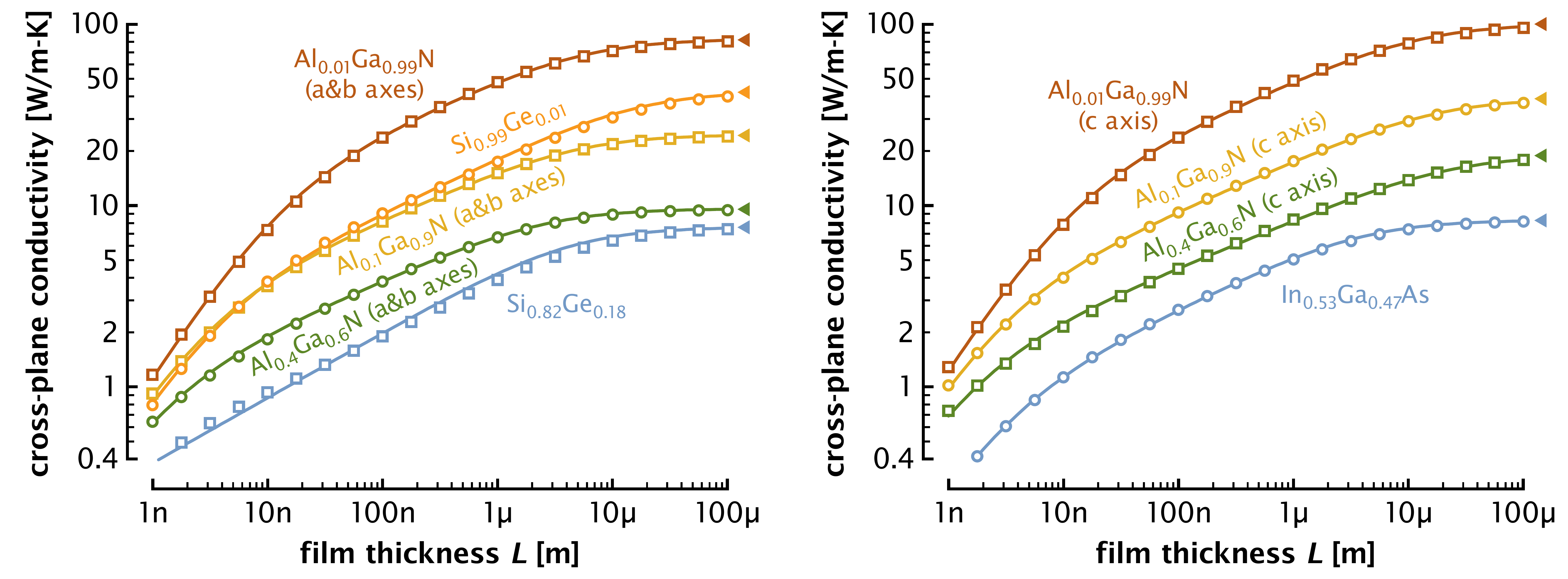}{(Color online.) Cross-plane film conductivity in alloys. Analytic curves per Eq. (\ref{parametricalloy}) with parameters listed in Table I fit \textit{ab-initio} values per Eq. (\ref{kappaL}) (symbols) within a few percent. The triangles mark the bulk conductivity.}
\myfigwide[!htb]{width=0.85\textwidth}{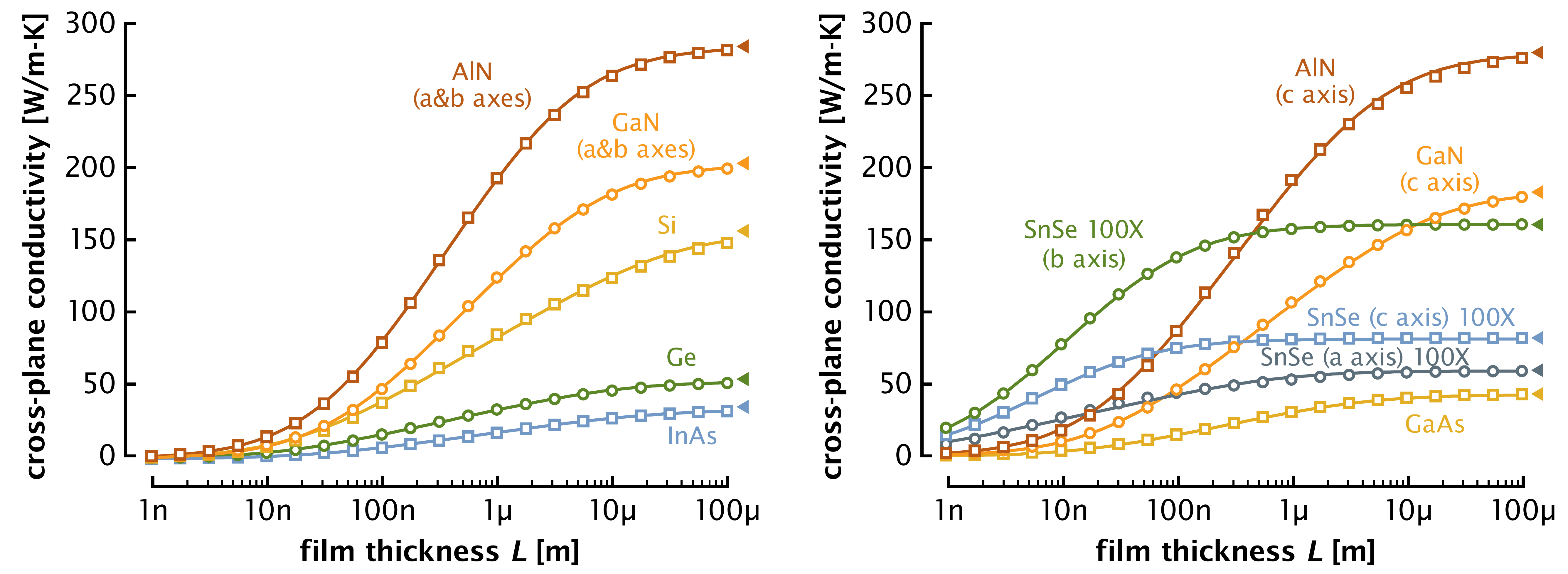}{(Color online.) Cross-plane film conductivity in single crystals. Analytic curves per Eq. (\ref{parametricsinglecrystal}) with parameters listed in Table I fit \textit{ab-initio} values per Eq. (\ref{kappaL}) (symbols) within a few percent. The triangles mark the bulk conductivity.}
\par
We now turn our attention to the case $n = 3$, which we have found to provide a reasonable description for \textit{ab-initio} scattering rates in single crystals (non-alloys). The primitive function of the integral in (\ref{kappaintegral}) is known analytically to be $\ln[u/(1+u)]$, prompting us to postulate
\begin{equation}
\text{non-alloys: } \kappa(L) \simeq \frac{\kappa_{\text{bulk}}}{\ln(L_{\text{d}}/L_{\text{b}})} \, \ln \left[ \frac{1+L/L_{\text{b}}}{1+L/L_{\text{d}}} \right] \label{parametricsinglecrystal}
\end{equation}
Appropriate diffusive $\kappa(L \gg L_{\text{d}}) \simeq \kappa_{\text{bulk}}$ and ballistic  $\kappa(L \ll L_{\text{b}}) \sim L$ limits are readily verified. In quasiballistic regime $L_{\text{b}} \ll L \ll L_{\text{d}}$ we find $\kappa \sim \ln(L/L_{\text{b}})$, which manifests itself as a quasilinear segment in the $\kappa(L)$ curves when plotted on semilogx scale (Fig. \ref{fig4-parametriccurves2}). Here too, the simple parametric form provides an accurate fit to the \textit{ab-initio} results for all considered compounds.
\npar%
In summary, we have analysed cross-plane thermal conductivity in thin semiconductor films from first principles. We observed, and explained, a quasiballistic conduction regime that displays a fractional-exponent and logarithmic thickness dependence in alloys and single crystals respectively. A simple suppression model, as well as two analytical parametric forms, are capable to match Monte Carlo solutions within a few percent regardless the compound's (an)isotropy. Film conductivities can be furthermore reliably reconstructed from MFP spectra, analogous to previously made observations for in-plane transport.
\npar%
This work was funded by \textsc{alma-bte}, a European Union Horizon 2020 project (Grant No. 645776) aimed at the development of an open-source multiscale BTE solver with industrial applications to power electronics.

\begin{thebibliography}{24}%
\makeatletter
\providecommand \@ifxundefined [1]{%
 \@ifx{#1\undefined}
}%
\providecommand \@ifnum [1]{%
 \ifnum #1\expandafter \@firstoftwo
 \else \expandafter \@secondoftwo
 \fi
}%
\providecommand \@ifx [1]{%
 \ifx #1\expandafter \@firstoftwo
 \else \expandafter \@secondoftwo
 \fi
}%
\providecommand \natexlab [1]{#1}%
\providecommand \enquote  [1]{``#1''}%
\providecommand \bibnamefont  [1]{#1}%
\providecommand \bibfnamefont [1]{#1}%
\providecommand \citenamefont [1]{#1}%
\providecommand \href@noop [0]{\@secondoftwo}%
\providecommand \href [0]{\begingroup \@sanitize@url \@href}%
\providecommand \@href[1]{\@@startlink{#1}\@@href}%
\providecommand \@@href[1]{\endgroup#1\@@endlink}%
\providecommand \@sanitize@url [0]{\catcode `\\12\catcode `\$12\catcode
  `\&12\catcode `\#12\catcode `\^12\catcode `\_12\catcode `\%12\relax}%
\providecommand \@@startlink[1]{}%
\providecommand \@@endlink[0]{}%
\providecommand \url  [0]{\begingroup\@sanitize@url \@url }%
\providecommand \@url [1]{\endgroup\@href {#1}{\urlprefix }}%
\providecommand \urlprefix  [0]{URL }%
\providecommand \Eprint [0]{\href }%
\providecommand \doibase [0]{http://dx.doi.org/}%
\providecommand \selectlanguage [0]{\@gobble}%
\providecommand \bibinfo  [0]{\@secondoftwo}%
\providecommand \bibfield  [0]{\@secondoftwo}%
\providecommand \translation [1]{[#1]}%
\providecommand \BibitemOpen [0]{}%
\providecommand \bibitemStop [0]{}%
\providecommand \bibitemNoStop [0]{.\EOS\space}%
\providecommand \EOS [0]{\spacefactor3000\relax}%
\providecommand \BibitemShut  [1]{\csname bibitem#1\endcsname}%
\let\auto@bib@innerbib\@empty
\bibitem [{\citenamefont {D.{G. Cahill}}\ \emph {et~al.}(2003)\citenamefont
  {D.{G. Cahill}}, \citenamefont {W.{K. Ford}}, \citenamefont {G.{D. Mahan}},
  \citenamefont {Majumdar}, \citenamefont {H.{J. Maris}}, \citenamefont
  {Merlin},\ and\ \citenamefont {S.{R. Phillpot}}}]{cahillreview1}%
  \BibitemOpen
  \bibfield  {author} {\bibinfo {author} {\bibnamefont {D.{G. Cahill}}},
  \bibinfo {author} {\bibnamefont {W.{K. Ford}}}, \bibinfo {author}
  {\bibnamefont {G.{D. Mahan}}}, \bibinfo {author} {\bibfnamefont
  {A.}~\bibnamefont {Majumdar}}, \bibinfo {author} {\bibnamefont {H.{J.
  Maris}}}, \bibinfo {author} {\bibfnamefont {R.}~\bibnamefont {Merlin}}, \
  and\ \bibinfo {author} {\bibnamefont {S.{R. Phillpot}}},\ }\href@noop {}
  {\bibfield  {journal} {\bibinfo  {journal} {J. Appl. Phys.}\ }\textbf
  {\bibinfo {volume} {91}},\ \bibinfo {pages} {793} (\bibinfo {year}
  {2003})}\BibitemShut {NoStop}%
\bibitem [{\citenamefont {D.{G. Cahill}}\ \emph {et~al.}(2014)\citenamefont
  {D.{G. Cahill}}, \citenamefont {P.{V. Braun}}, \citenamefont {Chen},
  \citenamefont {D.{R. Clarke}}, \citenamefont {Fan}, \citenamefont {K.{E.
  Godson}}, \citenamefont {Keblinski}, \citenamefont {W.{P. King}},
  \citenamefont {G.{D. Mahan}},\ and\ \citenamefont {{Majumdar, \textit{et
  al.}}}}]{cahillreview2}%
  \BibitemOpen
  \bibfield  {author} {\bibinfo {author} {\bibnamefont {D.{G. Cahill}}},
  \bibinfo {author} {\bibnamefont {P.{V. Braun}}}, \bibinfo {author}
  {\bibfnamefont {G.}~\bibnamefont {Chen}}, \bibinfo {author} {\bibnamefont
  {D.{R. Clarke}}}, \bibinfo {author} {\bibfnamefont {S.}~\bibnamefont {Fan}},
  \bibinfo {author} {\bibnamefont {K.{E. Godson}}}, \bibinfo {author}
  {\bibfnamefont {P.}~\bibnamefont {Keblinski}}, \bibinfo {author}
  {\bibnamefont {W.{P. King}}}, \bibinfo {author} {\bibnamefont {G.{D.
  Mahan}}}, \ and\ \bibinfo {author} {\bibfnamefont {A.}~\bibnamefont
  {{Majumdar, \textit{et al.}}}},\ }\href@noop {} {\bibfield  {journal}
  {\bibinfo  {journal} {Appl. Phys. Rev.}\ }\textbf {\bibinfo {volume} {1}},\
  \bibinfo {pages} {011305} (\bibinfo {year} {2014})}\BibitemShut {NoStop}%
\bibitem [{\citenamefont {Majumdar}(1993)}]{majumdar}%
  \BibitemOpen
  \bibfield  {author} {\bibinfo {author} {\bibfnamefont {A.}~\bibnamefont
  {Majumdar}},\ }\href@noop {} {\bibfield  {journal} {\bibinfo  {journal} {ASME
  J. Heat Trans.}\ }\textbf {\bibinfo {volume} {115}},\ \bibinfo {pages} {7}
  (\bibinfo {year} {1993})}\BibitemShut {NoStop}%
\bibitem [{\citenamefont {S.{V.J. Narumanchi}}\ \emph
  {et~al.}(2005)\citenamefont {S.{V.J. Narumanchi}}, \citenamefont {J.{Y.
  Murthy}},\ and\ \citenamefont {C.{H. Amon}}}]{extraref1}%
  \BibitemOpen
  \bibfield  {author} {\bibinfo {author} {\bibnamefont {S.{V.J. Narumanchi}}},
  \bibinfo {author} {\bibnamefont {J.{Y. Murthy}}}, \ and\ \bibinfo {author}
  {\bibnamefont {C.{H. Amon}}},\ }\href@noop {} {\bibfield  {journal} {\bibinfo
   {journal} {ASME J. Heat Trans.}\ }\textbf {\bibinfo {volume} {126}},\
  \bibinfo {pages} {946} (\bibinfo {year} {2005})}\BibitemShut {NoStop}%
\bibitem [{\citenamefont {Mittal}\ and\ \citenamefont
  {Mazumder}(2010)}]{extraref2}%
  \BibitemOpen
  \bibfield  {author} {\bibinfo {author} {\bibfnamefont {A.}~\bibnamefont
  {Mittal}}\ and\ \bibinfo {author} {\bibfnamefont {S.}~\bibnamefont
  {Mazumder}},\ }\href@noop {} {\bibfield  {journal} {\bibinfo  {journal} {ASME
  J. Heat Trans.}\ }\textbf {\bibinfo {volume} {132}},\ \bibinfo {pages}
  {052402} (\bibinfo {year} {2010})}\BibitemShut {NoStop}%
\bibitem [{\citenamefont {Maassen}\ and\ \citenamefont
  {Lundstrom}(2015)}]{extraref3}%
  \BibitemOpen
  \bibfield  {author} {\bibinfo {author} {\bibfnamefont {J.}~\bibnamefont
  {Maassen}}\ and\ \bibinfo {author} {\bibfnamefont {M.}~\bibnamefont
  {Lundstrom}},\ }\href@noop {} {\bibfield  {journal} {\bibinfo  {journal} {J.
  Appl. Phys.}\ }\textbf {\bibinfo {volume} {117}},\ \bibinfo {pages} {035104}
  (\bibinfo {year} {2015})}\BibitemShut {NoStop}%
\bibitem [{\citenamefont {Hua}\ and\ \citenamefont {A.{J.
  Minnich}}(2015)}]{minnichthinfilm}%
  \BibitemOpen
  \bibfield  {author} {\bibinfo {author} {\bibfnamefont {C.}~\bibnamefont
  {Hua}}\ and\ \bibinfo {author} {\bibnamefont {A.{J. Minnich}}},\ }\href@noop
  {} {\bibfield  {journal} {\bibinfo  {journal} {J. Appl. Phys.}\ }\textbf
  {\bibinfo {volume} {117}},\ \bibinfo {pages} {175306} (\bibinfo {year}
  {2015})}\BibitemShut {NoStop}%
\bibitem [{\citenamefont {Liu}\ \emph {et~al.}(2006)\citenamefont {Liu},
  \citenamefont {Etassam-Yazdani}, \citenamefont {Hussin},\ and\ \citenamefont
  {Asheghi}}]{SOIfilms}%
  \BibitemOpen
  \bibfield  {author} {\bibinfo {author} {\bibfnamefont {W.}~\bibnamefont
  {Liu}}, \bibinfo {author} {\bibfnamefont {K.}~\bibnamefont
  {Etassam-Yazdani}}, \bibinfo {author} {\bibfnamefont {R.}~\bibnamefont
  {Hussin}}, \ and\ \bibinfo {author} {\bibfnamefont {M.}~\bibnamefont
  {Asheghi}},\ }\href@noop {} {\bibfield  {journal} {\bibinfo  {journal} {IEEE
  Trans. Electr. Dev.}\ }\textbf {\bibinfo {volume} {53}},\ \bibinfo {pages}
  {1868} (\bibinfo {year} {2006})}\BibitemShut {NoStop}%
\bibitem [{\citenamefont {Liu}\ and\ \citenamefont {A.{A.
  Balandin}}(2005)}]{nitrides}%
  \BibitemOpen
  \bibfield  {author} {\bibinfo {author} {\bibfnamefont {W.}~\bibnamefont
  {Liu}}\ and\ \bibinfo {author} {\bibnamefont {A.{A. Balandin}}},\ }\href@noop
  {} {\bibfield  {journal} {\bibinfo  {journal} {J. Appl. Phys.}\ }\textbf
  {\bibinfo {volume} {97}},\ \bibinfo {pages} {073710} (\bibinfo {year}
  {2005})}\BibitemShut {NoStop}%
\bibitem [{\citenamefont {Yang}\ and\ \citenamefont
  {Dames}(2013)}]{kappacumulcurves}%
  \BibitemOpen
  \bibfield  {author} {\bibinfo {author} {\bibfnamefont {F.}~\bibnamefont
  {Yang}}\ and\ \bibinfo {author} {\bibfnamefont {C.}~\bibnamefont {Dames}},\
  }\href@noop {} {\bibfield  {journal} {\bibinfo  {journal} {Phys. Rev. B}\
  }\textbf {\bibinfo {volume} {87}},\ \bibinfo {pages} {035437} (\bibinfo
  {year} {2013})}\BibitemShut {NoStop}%
\bibitem [{\citenamefont {Li}\ \emph {et~al.}(2014)\citenamefont {Li},
  \citenamefont {Carrete}, \citenamefont {N.{A. Katcho}},\ and\ \citenamefont
  {Mingo}}]{ShengBTE_2014}%
  \BibitemOpen
  \bibfield  {author} {\bibinfo {author} {\bibfnamefont {W.}~\bibnamefont
  {Li}}, \bibinfo {author} {\bibfnamefont {J.}~\bibnamefont {Carrete}},
  \bibinfo {author} {\bibnamefont {N.{A. Katcho}}}, \ and\ \bibinfo {author}
  {\bibfnamefont {N.}~\bibnamefont {Mingo}},\ }\href@noop {} {\bibfield
  {journal} {\bibinfo  {journal} {Comp. Phys. Commun.}\ }\textbf {\bibinfo
  {volume} {185}},\ \bibinfo {pages} {1747} (\bibinfo {year}
  {2014})}\BibitemShut {NoStop}%
\bibitem [{\citenamefont {Mingo}\ \emph {et~al.}(2014)\citenamefont {Mingo},
  \citenamefont {Stewart}, \citenamefont {Broido}, \citenamefont {Lindsay},\
  and\ \citenamefont {Li}}]{abinitiobookchapter}%
  \BibitemOpen
  \bibfield  {author} {\bibinfo {author} {\bibfnamefont {N.}~\bibnamefont
  {Mingo}}, \bibinfo {author} {\bibfnamefont {D.}~\bibnamefont {Stewart}},
  \bibinfo {author} {\bibfnamefont {D.}~\bibnamefont {Broido}}, \bibinfo
  {author} {\bibfnamefont {L.}~\bibnamefont {Lindsay}}, \ and\ \bibinfo
  {author} {\bibfnamefont {W.}~\bibnamefont {Li}},\ }in\ \href@noop {} {\emph
  {\bibinfo {booktitle} {Length-Scale Dependent Phonon Interactions}}}\
  (\bibinfo  {publisher} {Springer},\ \bibinfo {year} {2014})\ pp.\ \bibinfo
  {pages} {137--173}\BibitemShut {NoStop}%
\bibitem [{\citenamefont {M.{S. Dresselhaus}}\ \emph
  {et~al.}(2007)\citenamefont {M.{S. Dresselhaus}}, \citenamefont {Chen},
  \citenamefont {M.{Y. Tang}}, \citenamefont {Yang}, \citenamefont {Lee},
  \citenamefont {Wang}, \citenamefont {Ren}, \citenamefont {Fleurial},\ and\
  \citenamefont {Gogna}}]{sige}%
  \BibitemOpen
  \bibfield  {author} {\bibinfo {author} {\bibnamefont {M.{S. Dresselhaus}}},
  \bibinfo {author} {\bibfnamefont {G.}~\bibnamefont {Chen}}, \bibinfo {author}
  {\bibnamefont {M.{Y. Tang}}}, \bibinfo {author} {\bibfnamefont
  {R.}~\bibnamefont {Yang}}, \bibinfo {author} {\bibfnamefont {H.}~\bibnamefont
  {Lee}}, \bibinfo {author} {\bibfnamefont {D.}~\bibnamefont {Wang}}, \bibinfo
  {author} {\bibfnamefont {Z.}~\bibnamefont {Ren}}, \bibinfo {author}
  {\bibfnamefont {J.-P.}\ \bibnamefont {Fleurial}}, \ and\ \bibinfo {author}
  {\bibfnamefont {P.}~\bibnamefont {Gogna}},\ }\href@noop {} {\bibfield
  {journal} {\bibinfo  {journal} {Adv. Mater.}\ }\textbf {\bibinfo {volume}
  {19}},\ \bibinfo {pages} {1043} (\bibinfo {year} {2007})}\BibitemShut
  {NoStop}%
\bibitem [{\citenamefont {J.{A. del Alamo}}(2011)}]{arsenides}%
  \BibitemOpen
  \bibfield  {author} {\bibinfo {author} {\bibnamefont {J.{A. del Alamo}}},\
  }\href@noop {} {\bibfield  {journal} {\bibinfo  {journal} {Nature}\ }\textbf
  {\bibinfo {volume} {479}},\ \bibinfo {pages} {317} (\bibinfo {year}
  {2011})}\BibitemShut {NoStop}%
\bibitem [{\citenamefont {L.{-D. Zhao}}\ \emph {et~al.}(2014)\citenamefont
  {L.{-D. Zhao}}, \citenamefont {S.{-H. Lo}}, \citenamefont {Zhang},
  \citenamefont {Sun}, \citenamefont {Tan}, \citenamefont {Uher}, \citenamefont
  {Wolverton}, \citenamefont {V.{P. Dravid}},\ and\ \citenamefont {M.{G.
  Kanatzidis}}}]{snsenature}%
  \BibitemOpen
  \bibfield  {author} {\bibinfo {author} {\bibnamefont {L.{-D. Zhao}}},
  \bibinfo {author} {\bibnamefont {S.{-H. Lo}}}, \bibinfo {author}
  {\bibfnamefont {Y.}~\bibnamefont {Zhang}}, \bibinfo {author} {\bibfnamefont
  {H.}~\bibnamefont {Sun}}, \bibinfo {author} {\bibfnamefont {G.}~\bibnamefont
  {Tan}}, \bibinfo {author} {\bibfnamefont {C.}~\bibnamefont {Uher}}, \bibinfo
  {author} {\bibfnamefont {C.}~\bibnamefont {Wolverton}}, \bibinfo {author}
  {\bibnamefont {V.{P. Dravid}}}, \ and\ \bibinfo {author} {\bibnamefont {M.{G.
  Kanatzidis}}},\ }\href@noop {} {\bibfield  {journal} {\bibinfo  {journal}
  {Nature}\ }\textbf {\bibinfo {volume} {508}},\ \bibinfo {pages} {373}
  (\bibinfo {year} {2014})}\BibitemShut {NoStop}%
\bibitem [{\citenamefont {Carrete}\ \emph {et~al.}(2014)\citenamefont
  {Carrete}, \citenamefont {Mingo},\ and\ \citenamefont {Curtarolo}}]{snse}%
  \BibitemOpen
  \bibfield  {author} {\bibinfo {author} {\bibfnamefont {J.}~\bibnamefont
  {Carrete}}, \bibinfo {author} {\bibfnamefont {N.}~\bibnamefont {Mingo}}, \
  and\ \bibinfo {author} {\bibfnamefont {S.}~\bibnamefont {Curtarolo}},\
  }\href@noop {} {\bibfield  {journal} {\bibinfo  {journal} {Appl. Phys.
  Lett.}\ }\textbf {\bibinfo {volume} {105}},\ \bibinfo {pages} {101907}
  (\bibinfo {year} {2014})}\BibitemShut {NoStop}%
\bibitem [{\citenamefont {Vermeersch}\ \emph
  {et~al.}(2015{\natexlab{a}})\citenamefont {Vermeersch}, \citenamefont
  {Carrete}, \citenamefont {Mingo},\ and\ \citenamefont {Shakouri}}]{levy1}%
  \BibitemOpen
  \bibfield  {author} {\bibinfo {author} {\bibfnamefont {B.}~\bibnamefont
  {Vermeersch}}, \bibinfo {author} {\bibfnamefont {J.}~\bibnamefont {Carrete}},
  \bibinfo {author} {\bibfnamefont {N.}~\bibnamefont {Mingo}}, \ and\ \bibinfo
  {author} {\bibfnamefont {A.}~\bibnamefont {Shakouri}},\ }\href@noop {}
  {\bibfield  {journal} {\bibinfo  {journal} {Phys. Rev. B}\ }\textbf {\bibinfo
  {volume} {91}},\ \bibinfo {pages} {085202} (\bibinfo {year}
  {2015}{\natexlab{a}})}\BibitemShut {NoStop}%
\bibitem [{\citenamefont {Li}\ \emph {et~al.}(2012)\citenamefont {Li},
  \citenamefont {Lindsay}, \citenamefont {D.{A. Broido}}, \citenamefont {D.{A.
  Stewart}},\ and\ \citenamefont {Mingo}}]{MgSiSn}%
  \BibitemOpen
  \bibfield  {author} {\bibinfo {author} {\bibfnamefont {W.}~\bibnamefont
  {Li}}, \bibinfo {author} {\bibfnamefont {L.}~\bibnamefont {Lindsay}},
  \bibinfo {author} {\bibnamefont {D.{A. Broido}}}, \bibinfo {author}
  {\bibnamefont {D.{A. Stewart}}}, \ and\ \bibinfo {author} {\bibfnamefont
  {N.}~\bibnamefont {Mingo}},\ }\href@noop {} {\bibfield  {journal} {\bibinfo
  {journal} {Phys. Rev. B}\ }\textbf {\bibinfo {volume} {86}},\ \bibinfo
  {pages} {174307} (\bibinfo {year} {2012})}\BibitemShut {NoStop}%
\bibitem [{\citenamefont {N.{A. Katcho}}\ \emph {et~al.}(2012)\citenamefont
  {N.{A. Katcho}}, \citenamefont {Mingo},\ and\ \citenamefont {D.{A.
  Broido}}}]{katcho_lattice_2012}%
  \BibitemOpen
  \bibfield  {author} {\bibinfo {author} {\bibnamefont {N.{A. Katcho}}},
  \bibinfo {author} {\bibfnamefont {N.}~\bibnamefont {Mingo}}, \ and\ \bibinfo
  {author} {\bibnamefont {D.{A. Broido}}},\ }\href@noop {} {\bibfield
  {journal} {\bibinfo  {journal} {Phys. Rev. B}\ }\textbf {\bibinfo {volume}
  {85}},\ \bibinfo {pages} {115208} (\bibinfo {year} {2012})}\BibitemShut
  {NoStop}%
\bibitem [{\citenamefont {E.{N. Economou}}(1983)}]{economou}%
  \BibitemOpen
  \bibfield  {author} {\bibinfo {author} {\bibnamefont {E.{N. Economou}}},\
  }\href@noop {} {\emph {\bibinfo {title} {Green's Functions in Quantum
  Physics}}},\ \bibinfo {edition} {2nd}\ ed.\ (\bibinfo  {publisher}
  {Springer},\ \bibinfo {address} {Berlin},\ \bibinfo {year}
  {1983})\BibitemShut {NoStop}%
\bibitem [{\citenamefont {J.{-P.M. P\'eraud}}\ and\ \citenamefont {N.{G.
  Hadjiconstantinou}}(2011)}]{VRMC}%
  \BibitemOpen
  \bibfield  {author} {\bibinfo {author} {\bibnamefont {J.{-P.M. P\'eraud}}}\
  and\ \bibinfo {author} {\bibnamefont {N.{G. Hadjiconstantinou}}},\
  }\href@noop {} {\bibfield  {journal} {\bibinfo  {journal} {Phys. Rev. B}\
  }\textbf {\bibinfo {volume} {84}},\ \bibinfo {pages} {205331} (\bibinfo
  {year} {2011})}\BibitemShut {NoStop}%
\bibitem [{sup()}]{supplmat}%
  \BibitemOpen
  \href@noop {} {}\bibinfo {howpublished} {See supplemental material at [URL to
  be inserted by AIP] for tabulated cumulative conductivity
  functions}\BibitemShut {NoStop}%
\bibitem [{\citenamefont {Vermeersch}\ \emph
  {et~al.}(2015{\natexlab{b}})\citenamefont {Vermeersch}, \citenamefont
  {A.{M.S. Mohammed}}, \citenamefont {Pernot}, \citenamefont {Y.{R. Koh}},\
  and\ \citenamefont {Shakouri}}]{levy2}%
  \BibitemOpen
  \bibfield  {author} {\bibinfo {author} {\bibfnamefont {B.}~\bibnamefont
  {Vermeersch}}, \bibinfo {author} {\bibnamefont {A.{M.S. Mohammed}}}, \bibinfo
  {author} {\bibfnamefont {G.}~\bibnamefont {Pernot}}, \bibinfo {author}
  {\bibnamefont {Y.{R. Koh}}}, \ and\ \bibinfo {author} {\bibfnamefont
  {A.}~\bibnamefont {Shakouri}},\ }\href@noop {} {\bibfield  {journal}
  {\bibinfo  {journal} {Phys. Rev. B}\ }\textbf {\bibinfo {volume} {91}},\
  \bibinfo {pages} {085203} (\bibinfo {year} {2015}{\natexlab{b}})}\BibitemShut
  {NoStop}%
\bibitem [{\citenamefont {{Abs da Cruz}}\ \emph {et~al.}(2013)\citenamefont
  {{Abs da Cruz}}, \citenamefont {N.{A. Katcho}}, \citenamefont {Mingo},\ and\
  \citenamefont {R.{G.A Veiga}}}]{fractionallaw}%
  \BibitemOpen
  \bibfield  {author} {\bibinfo {author} {\bibfnamefont {C.}~\bibnamefont {{Abs
  da Cruz}}}, \bibinfo {author} {\bibnamefont {N.{A. Katcho}}}, \bibinfo
  {author} {\bibfnamefont {N.}~\bibnamefont {Mingo}}, \ and\ \bibinfo {author}
  {\bibnamefont {R.{G.A Veiga}}},\ }\href@noop {} {\bibfield  {journal}
  {\bibinfo  {journal} {J. Appl. Phys.}\ }\textbf {\bibinfo {volume} {114}},\
  \bibinfo {pages} {164310} (\bibinfo {year} {2013})}\BibitemShut {NoStop}%
\end{thebibliography}
%
\end{document}